\documentclass[10pt]{article}

\usepackage{amsmath}
\usepackage{amssymb}

\usepackage{graphicx}

\usepackage{cite}

\usepackage{color} 

\usepackage[labelfont=bf,labelsep=period,justification=raggedright]{caption}

\makeatletter
\renewcommand{\@biblabel}[1]{\quad#1.}
\makeatother

\date{}

\begin{document}

\begin{flushleft}
{\Large
\textbf{How should prey animals respond to uncertain threats?}
}
\\
Joel Zylberberg$^{1,2,\ast}$, 
Michael Robert DeWeese$^{1,2,3}$, 
\\
\bf{1} Department of Physics, University of California, Berkeley, CA, USA
\\
\bf{2} Redwood Center for Theoretical Neuroscience, University of California, Berkeley, CA, USA
\\
\bf{3} Helen Wills Neuroscience Institute, University of California, Berkeley, CA, USA
\\
$\ast$ E-mail: joelz@berkeley.edu
\end{flushleft}

\section*{Abstract}

A prey animal surveying its environment must decide whether there is a dangerous predator present or not. If there is, it may flee. Flight has an associated cost, so the animal should not flee if there is no danger. However, the prey animal cannot know the state of its environment with certainty, and is thus bound to make some errors.  We formulate a probabilistic automaton model of a prey animal's life and use it to compute the optimal escape decision strategy, subject to the animal's uncertainty. The uncertainty is a major factor in determining the decision strategy: only in the presence of uncertainty do economic factors (like mating opportunities lost due to flight) influence the decision. We performed computer simulations and found that \emph{in silico} populations of animals subject to predation evolve to display the strategies predicted by our model, confirming our choice of objective function for our analytic calculations. To the best of our knowledge, this is the first theoretical study of escape decisions to incorporate the effects of uncertainty, and to demonstrate the correctness of the objective function used in the model.

\section{Introduction}
Prey animals frequently assess their surroundings to identify potential threats to their safety. If an animal does not flee soon enough in the presence of a predator (type I error), it may be injured or killed. If it flees when there is no legitimate threat (type II error), it wastes metabolic energy and loses mating or foraging opportunities~\cite{Nelson, Creswell}. However, predators can be camouflaged and prey animals have limited sensory systems, so escape decisions must often be made with imperfect information. 

Previous studies have not investigated how escape decisions might be affected by prey animals' degree of certainty about their environment. Indeed, the predominant assumption in the field appears to be that this uncertainty is not important, and that so long as the prey animal knows the most likely state of the environment (or the expected value of the state), they can still make economically optimal decisions. We question this assumption.

 We explicitly consider the animal's uncertainty in our model and subsequently demonstrate that, when a prey animal knows the environmental state with certainty, the optimal decision strategy is simply to flee whenever a threat is present. This strategy is independent of any ``economic" factors -- predator lethality, predator frequency, loss of mating opportunities, etc. When the state of the environment is less certain, the animal is bound to make errors, and the optimal balance between type I and II errors is determined by economic factors. This is in contrast with previous theoretical studies  (\cite{YdenDill, Cooper2007, Cooper2010, Cooper2006, Broom2005}, and others), which have assumed that prey animals have perfect knowledge of their surroundings (or, equivalently, that the uncertainty is unimportant, as discussed above) and that their decisions are made on purely economic grounds. Our result suggests that uncertainty may play a key role in making economics relevant in decision making.

It has been observed that prey animals increase their flight initiation distance (predator-prey distance at which they flee) when intruders begin their approach from farther away\cite{Blumstein}. This observation challenges strictly economic decision models (although Cooper\cite{CooperINT} has modeled this by assuming the prey animal has multiple ``risk" functions and chooses one based on the predator behavior, which includes starting distance) but may have a simple interpretation: when intruders approach from farther away, the animal has longer to detect and assess the threat, thus fleeing sooner. This ``increased information" interpretation is also consistent with the observations that prey animals increase their flight initiation distance when an intruder has a higher approach speed~\cite{Cooper2006}, and that odors~\cite{Apfelbach,Ylonen} and shapes~\cite{hemmi2009} are important in eliciting defensive behaviors -- the approach speed may indicate that the intruder is likely to attack, while the odor and morphology may indicate that the intruder is a potentially dangerous one.

The current models of escape decisions (\cite{YdenDill, Cooper2007, Cooper2010, Cooper2006, Broom2005}, and others) have yet to incorporate factors such as  olfaction and intruder morphology into their models. Such factors are difficult to include in economic models. For example, what is the cost associated with a particular smell? 

Inspired by these observations, we propose a new approach for studying escape decisions in prey animals, namely that they are engaged in a decision-theoretic process, wherein they must decide, with imperfect information, whether the current environment is likely to pose a significant enough threat to their safety that they should flee. This view is supported by observations of active risk assessment behaviors in prey animals\cite{Hemmi2010, Schaik, Creswell09}.

Sih~\cite{Sih} has studied the decision to re-emerge from a burrow after flight, when the animal does not know if the predator is still present. Sih's work is the closest in spirit to the current study, but it does not address how the initial escape decision is affected by uncertainty. No previous study satisfactorily addresses the issue of determining what objective function, when maximized, accurately predicts the escape decision strategy that is selected by evolution. 

We demonstrate through computer simulation that animals subject to predation naturally evolve to display the strategy predicted by our model, confirming our choice of objective function. 

\section{Methods and Models}

\subsection{Analytical calculations} 

As a starting point, we will assume that the prey animal chooses the strategy $S$ that maximizes its genetic contribution to subsequent generations (see\cite{Janetos1981, Parker} for criticism and discussion of optimality models), defined by the intrinsic rate of increase in prevalence of strategy $S$ in the population~\cite{Parker,hairston}; $r(S)=[N(S)^{-1}] d N(S)/dt$, where $N(S)$ is the number of animals adopting strategy $S$, and $t$ is time (note that we use the symbol $r$ for the same quantity that Hairston \emph{et al.}~\cite{hairston} call $m$). Since the large $r$ strategies grow more quickly in terms of number of adherents, the population should evolve towards the strategy that maximizes $r$. We later verify this assumption. We stress that our objective function is \emph{rate of reproduction} and not ``survival of the fittest". A genotype that  leads to long-lived animals, who fail to reproduce, is unlikely to significantly increase in prevalence over time. 

We formulate a probabilistic automaton~\cite{rabin} model of the life of a prey animal (Fig.~1) to compute $r(S)$. The model operates in discrete time, thus we are using the approximation $r = [N(S)^{-1}] dN(S)/dt \approx [N(S)^{-1}] \Delta N(S)/ \Delta t$. In every time step, every animal follows a complete path through the graph, beginning in the starting state (``start"), and ending either back at the start, or in death.

In each time step, the animal assesses a potential threat. For concreteness, we imagine the animal asking ``Is that object likely to try to kill me?" Animals that do not flee from a real threat may be killed by a predator, while those that do flee, escape. Those animals that are not killed by predators may mate, and they may or may not die of causes other than predation. 

We group potential threats into discrete ``zones" in predator-prey distance; Fig.~1 illustrates an example with four such zones. The model can utilize continuous distances without affecting our conclusions. The object the animal assesses may or may not be a real threat -- the animal does not have access to the ground truth. We explicitly incorporate this uncertainty by assigning probability $p_i$ of correct threat detection in zone $i$, leading to flight, and probability $q_i$ of a false positive in zone $i$, leading to flight; $q_i$ and $p_i$ are related by the receiver-operator-characteristic (ROC) curve ($p_i = f_i(q_i)$; see Fig.~2).

Qualitatively, our automaton model captures many features pertinent to real prey animals. Effects like periodicity of mating opportunities and threat frequency, maturation periods, learning during the lifetime of the animal~\cite{hemmi2009,rao}, and sexual reproduction (as opposed to asexual), are omitted in the interest of simplicity, but our automaton could be amended to incorporate these considerations. We have confirmed with a computer simulation that our results are unchanged when the animals undergo sexual, rather than asexual reproduction (results not shown). 

Figure~2 presents an example ROC curve, for the case in which the animal makes its choice based on a single, scalar parameter (the ``score"). The problem of choosing an escape strategy amounts to choosing where on the ROC curve the decision rule should lie. It can be specified in each zone either by $p_i$, $q_i$, or a threshold $\tau_i$ (see Fig.~2). 

To simplify our notation, we first define the variables $a,b,c$ to be the probabilities, in a given time step, of being killed by a predator, of not fleeing and not being killed by a predator, and of fleeing, respectively. By tracing paths in Fig.~1, we find that

\begin{eqnarray}
a& = &  \sum_{i}   \delta_i  \alpha \left[1-p_{i} \right]L_i     \\
b& = &     \sum_{i}   \delta_i   \left(  \alpha \left[1-p_i \right] \left[1-L_i \right] + \left[1-\alpha \right] \left[1-q_i \right]   \right)  \nonumber \\
c& = &     \sum_{i}   \delta_i  \left(   \alpha p_i + \left[1-\alpha \right] q_i  \right) \nonumber \\
1 &=& a + b+ c.    \nonumber \\
1  &=& \sum_i \delta_i \nonumber
\label{eq:abc}
\end{eqnarray}

where $\alpha$ is the probability that a given object is actually a threat, and $L_i$ is the probability that failure to immediately flee a predator, initally in distance zone $i$, will be lethal.

The expectation value of the objective function ($r = [N(S)^{-1}] dN(S)/dt \approx [N(S)^{-1}] \Delta N(S)/\Delta t$) is given by multiplying the result of a given outcome by the probability of that outcome, and summing over all possible outcomes.  These outcomes are as follows: animals that do not flee will mate with probability $m$, producing $n$ progeny, while those that do flee will suffer a reduced mating rate $m(1-R)$, producing $n$ progeny. Animals that die are removed from the population, thus the value of this outcome is $-1$. Those animals that are not already killed by the predator die with probability $\sigma$, which again has a value $-1$. Thus, our expectation value is

\begin{eqnarray}
E[r(\{q_i\})]  & \approx & b(\{q_i\})mn + c(\{q_i\})mn \left[1-R \right] -a(\{q_i\}) - \sigma (1-a\{q_i\}) \\ \nonumber
& \approx &  \left[ 1-a(\{q_i\}) \right] mn - c(\{q_i\})Rmn - a(\{q_i\}) - \sigma .  
\label{eq:objective}
\end{eqnarray}

The second line follows from the first since  $a + b + c = 1$, and, since both $\sigma$ and $a$ are expected to be small, the product $\sigma a$ can safely be ignored. 

Since the animals in our model assess one potential threat per unit time, the size of the ``time steps" in our model is fairly short (seconds, or possibly minutes). In the real world, we expect that actual threats are relatively uncommon (for example, the probability of encountering a real threat in any given short time period is small): $\alpha$ should be a relatively small quantity. Thus, $a$ is small for real prey animals. Furthermore, since the time steps are fairly short, the probability $\sigma$ of dying from starvation or disease in any time step is quite small. Thus, our $\sigma a <<1$ approximation (above) is reasonable.

The anticipated escape response threshold maximizes the expectation value of the objective function $E[r(\{q_i\})]$, subject to the constraints $p_i = f_i(q_i)$ imposed by the ROC curves for each zone. 

In the standard fashion~\cite{Boas}, we utilize the method of Lagrange multipliers by defining a Lagrange function $\Omega = E[r] + \sum_i \xi_i (p_i-f_i(q_i))$. The set $\{\xi_i\}$, then, is then our set of (unknown) Lagrange multipliers, and we optimize by solving (for all $i$)

\begin{eqnarray}
\frac{\partial \Omega}{\partial p_i} & = & \frac{\partial E[r(\{q_i\})]}{\partial p_i} + \xi_i = 0 \\
\frac{\partial \Omega}{\partial q_i}  & = & \frac{\partial E[r(\{q_i\})]}{\partial q_i} - \xi f_i'(q_i) = 0  \nonumber\\
\frac{\partial \Omega}{\partial \xi_i}  &=&  p_i- f_i(q_i) =0.  \nonumber
\label{eq:optimize}
\end{eqnarray}

The last of these equations enforces the constraint. The first two equations yield

\begin{eqnarray}
\xi_i & = & -\alpha L_i \left[1 + mn \right] + \alpha Rmn \\
\xi_i f_i'(q_i) &=& - Rmn \left[ 1-\alpha \right] .\nonumber 
\label{eq:optimize3}
\end{eqnarray}

And the solution to our optimization problem is (for $\xi_i \ne 0$)

\begin{eqnarray}
f_i'(q_i) &=& \frac{Rmn \left[ 1-\alpha \right] }{ \alpha \left( mn[L_i - R] + L_i \right)    }\\
p_i & =& f_i(q_i) . \nonumber 
\label{eq:solution}
\end{eqnarray}

Note that, were all of the threats in one distance zone, Eq.~5 still yields the optimal result. Thus, the globally optimal solution consists of making the locally optimal decision for each zone, as one might expect.

For an explicit computation of where the decision threshold should lie, we require information about the ROC curve. As an example, we assume that the score is distributed as $g_i(z | danger) = \mathcal{N}(0,1)$ in the presence of danger in zone $i$, where $\mathcal{N}(\mu,\Sigma)$ represents a Gaussian (or normal) distribution with mean $\mu$ and standard deviation $\Sigma$. Now let the distribution of scores in the absence of danger in zone $i$ be $g_i(z|no~danger) = \mathcal{N}(\omega_i,1)$ for some $\omega_i<0$. 

Note that, given a Gaussian distributed variable $y$ with arbitrary mean $\mu$ and variance $\Sigma$, we can choose to operate on the variable $x = (y - \mu) / \Sigma$, which will be distributed as $\mathcal{N}(0,1)$. Thus, within the realm of Gaussian distributed scores, we are losing no generality by considering $g_i(z | danger) = \mathcal{N}(0,1)$. Choosing the variance of ``no threat" score distribution to be the same as that of the distribution conditioned on presence of danger does entail a loss of generality, but it simplifies the analysis greatly and thus we do it for the purposes of this example. Given the distributions, we can define the values $(p_i,q_i)$ as a function of the decision threshold $\tau_i$. Let the animal decide that it is in danger for $z > \tau_i$, and that it is not for $z \leq \tau_i$. Then

\begin{eqnarray}
p_i &=& f_i(q_i(\tau_i)) = \int_{\tau_i}^\infty \frac{dz}{\sqrt{2 \pi}} e^{-z^2/2} \\
q_i(\tau_i) & = & \int_{\tau_i}^{\infty}  \frac{dz}{\sqrt{2 \pi }} e^{-(z-\omega_i)^2/2}.     \nonumber 
\label{ROC_ex}
\end{eqnarray}

We need the derivative $f'_i(q_i)$ to implement the results of our optimization calculation. Using the chain rule, 

\begin{eqnarray}
f'_i(q_i(\tau_i))& = &\frac{d p_i}{d \tau_i} \frac{ d \tau_i}{d q_i} \\
& = & \exp\left( \frac{\omega_i^2 - 2\omega_i \tau_i}{2} \right). \nonumber
\label{chain_rule}
\end{eqnarray}

Therefore, the optimal threshold for the $i^{th}$ zone is 

\begin{eqnarray}
\tau_i &=& \frac{\omega_i}{2} - \frac{1}{\omega_i} \ln \left( \frac{Rmn(1-\alpha)}{\alpha \left[ mn(L_i-R) + L_i\right] } \right). 
\label{optimum_ex}
\end{eqnarray}

It is clear that as $|\omega_i|$ increases (more obvious threats and thus less uncertainty), the second term, which contains all of the economic factors about the environment, becomes less important in determining the decision threshold.

The decreasing importance of the ``economic" term with increasing $|\omega|$ is not true for all possible score distributions. We have demonstrated that this conclusion does apply to Gaussian distributions. Indeed, it also applies to any unimodal distribution in the exponential family $g(z) \propto e^{-\lambda |z|^{\nu}}$ with even $\nu \geq 2$. 

This can be seen by noting that, if the distribution of scores in the presence of a threat is $g(z|threat) \propto e^{-\lambda |z|^\nu}$, and in the absence of a threat is $g(z|no~threat) \propto e^{- \lambda |z-\omega|^\nu}$, then the derivative of the ROC curve $f'(q)$ is given by

\begin{eqnarray}
f'(q(\tau))=  \exp\left( \lambda |\tau|^\nu - \lambda|\tau - \omega|^\nu\right) \\
\rightarrow \lambda^{-1} \ln[f'(q)] = |\tau|^\nu - |\tau - \omega|^\nu.  \nonumber
\label{gen_dist}
\end{eqnarray}

Solving for $\tau$ is hard for general $\nu$. Consider, for example the case where $\tau \geq 0$ and $\tau \geq \omega$. Then we see that $\lambda^{-1} \ln[f'(q)] = \tau^\nu - (\tau - \omega)^\nu$. For $\nu = 1$, this yields no solution for $\tau$ because the derivative of the ROC curve is independent of $\tau$. This is peculiar to the exponential distribution, which is a pathological case in this sense. 

For even $\nu$, with no restrictions on $\tau$, we see that $\lambda^{-1} \ln[f'(q)] = \tau^\nu - (\tau - \omega)^\nu$ (the absolute value signs disappear for even $\nu$). Expanding $(\tau - \omega)^\nu$ using binomial theorem, we find that $\lambda^{-1} \ln[f'(q)] = \tau^\nu - \tau^\nu - \sum_{j=1}^{\nu} {\nu\choose j} \tau^{\nu-j} (-\omega)^j$. Now, the $\tau^\nu$ terms cancel, and we can divide through by one power of $\omega$, yielding $\lambda^{-1} \omega^{-1} \ln[f'(q)] = \sum_{j=1}^{\nu} {\nu\choose j} \tau^{\nu-j} (-\omega)^{j-1}$. As in the Gaussian ($\nu = 2$) case, we see that increasing $|\omega|$ de-weights the economic $f'(q)$ term.

Thus, we can be assured that the decreasing importance of the economic term with increasing $|\omega|$ is true for all unimodal exponential distributions of the form $g(z) \propto e^{-\lambda |z|^{\nu}}$ with even $\nu \geq 2$. For $\nu>1$ and values of $\nu$ that are not even integers, there are some regimes in which the leading-order terms in $\tau$ still cancel, however it is difficult to prove that our result holds in the most general case.

We note that, while it simplified our automaton model and our notation, nowhere was it necessary to assume that the danger occurs in discrete zones in distance. One could instead utilize a continuous distance measure by considering an infinitely large set of possible values of $i$, with each one corresponding to a particular point in space.

\subsection{Simulation Experiments}
To verify that our objective function is the one selected for by natural evolution, we perform a computer simulation of a population of prey animals subject to predation.

Our simulation contains a population of animals whose life cycles are described by the probabilistic automaton model (Fig.~1). At each time step of the simulation, the animals are considered one-by-one.  A pseudo-random number generator determines whether a prey animal will see a real threat (with probability $\alpha$) or not. The threats are all in the same distance zone, since this simplifies the simulation, and we have shown that the optimal solution for many zones is to use the locally optimal solution in each separate zone (Eq.~5).

The animal is then presented with a ``score" variable, with which it makes its decision. As in our analytic example, the scores are randomly drawn from the $\mathcal{N}(0,1)$ distribution if the threat is real, or from the $\mathcal{N}(\omega,1)$ distribution if the threat is fake. If the ``score" is above the animal's threshold, it chooses to flee. Otherwise it does not. The determination of which animals get to mate, or get killed by a predator is also done with a pseudo-random number generator, and follows the description in Fig.~1.

Those animals that do mate produce $n$ progeny. Each offspring has a decision threshold that is equal to its parent's, plus Gaussian noise of mean zero and fixed (small) standard deviation. This variation allows the population to explore the strategy space. The population in our simulation thus has the two key features (heritability, and variability) that allow for evolution.

At the end of every time step, the population is trimmed so that it does not get too large. This is done by killing random individuals, thus inducing no selection pressure. This is represented by the value $\sigma$ in our automaton model.

We initialize the simulation with a population of animals whose decision thresholds are drawn from a uniform distribution.

\section{Results and Discussion}

\subsection{The optimal decision strategy depends on the environment and varies with the animal's uncertainty about the state of the environment}

The strategy that maximizes the expectation value of $r$, subject to the $p_i = f_i(q_i)$ constraints imposed by the ROC curves, is given by (for all zones $i$)

\begin{equation}
f'_i(q_i) = \frac{Rmn(1-\alpha)}{\alpha \left[ (L_{i} - R)mn + L_{i} \right]}.
\label{eq:optimalfprime}
\end{equation}

Fig.~2 shows that low thresholds yield small derivatives $f'_i(q_i)$ (for low thresholds, a small increase in $\tau$ increases $q$ much more than $p$), and vice versa. We observe that increased predator density and lethality leads to more timid prey animals (low $\tau$), while increased reproductive flight cost $Rmn$ leads to bolder ones.

To investigate the influence of uncertainty on the decision strategy, we consider a specific example: the decision is made based on a single ``score" variable $z$, which is distributed as $g_i(z | danger) = \mathcal{N}(0,1)$ in the presence of danger in zone $i$, where $\mathcal{N}(a,b)$ is a Gaussian (or normal) distribution with mean $a$ and standard deviation $b$.

This score may be, for example, the output of a neural network that takes into account all of the information available to the animal, including information about the predator behavior, odor, morphology, etc. The use of a single score for the decision can be understood as a dimensionality reduction step: the high-dimensional sensory data is reduced to a single scalar value, upon which the decision can be based. In the case of an animal with a ``command neuron" ({\em e.g.}, the Mauthner cell)~\cite{Rock, Roberts, Zottoli, Korn}, the score we refer to may be related to the membrane potential, which is a function of all the synaptic inputs to that cell from the sensory processing network.  

Let the distribution of scores in the absence of danger in zone $i$ be $g_i(z|no~danger) = \mathcal{N}(\omega_i,1)$ for some $\omega_i <0$. The absolute value of $\omega_i$ defines the reliability of the information available to the animal: larger $|\omega_i|$ implies more reliable information. Let the animal decide that it is in danger  for $z > \tau_i$, and that it is not in danger for $z  \leq \tau_i$. The optimal threshold $\tau_i$ for zone $i$ is 

\begin{equation}
\tau_i = \frac{\omega_i}{2} - \frac{1}{\omega_i} \ln[f'_i(q_i)],  
\label{eq:optimalex}
\end{equation}

and $f'_i(q_i)$, which incorporates all of the economic factors in the probabilistic automaton model, is given by Eq.~5.

For large $|\omega_i|$, the dependence on the logarithmic term is small and, for $f'_i(q_i) >0$ (which is true, for example, when $L_i>R$ and $\alpha \neq 1$), the threshold is $\tau_i \approx \omega_i /2$. This strategy resembles a maximum likelihood estimator (MLE), which, in this example, would be given by $\tau_i = \omega_i/2$. As $| \omega_i|$ decreases, the economic factors become more important in determining the threshold $\tau_i$. This conclusion (demonstrated in Fig.~3) is independent of the details of our probabilistic automaton model. The specifics of the automaton determine the optimal $f'_i(q_i)$, but Eq.~\ref{eq:optimalex} shows us that, independent of $f'_i(q_i)$, the strategy still changes from maximum likelihood to economic cost reduction, as the amount of uncertainty in the information increases. 

This result is true for Gaussian-distributed score variables, but is not true for all distributions. However, it is straightforward to prove that the result holds for all unimodal distributions in the exponential family $g(z) \propto  e^{-\lambda |z|^{\nu}}$ for even $\nu \geq 2$. According to the central limit theorem, most variables that are weighted averages of many random components are Gaussian distributed. Thus, our conclusion is likely to be applicable to many real-world examples.

\subsection{An \emph{in silico} population of prey animals evolves to display the decision strategy predicted by our analytical calculations}

To verify our choice of objective function, and the approximations made in our calculation, we performed a computer simulation of a population of prey animals subject to predation. Unlike previous work\cite{Floreano}, we did not define an objective function in our simulation: the animals in our simulation had no indication of what we thought they should be accomplishing with their escape strategy. They simply mated, died of predation, and were killed by non-predation-related causes. Those animals that did mate produced children whose escape thresholds were copies of the parent's threshold, with added Gaussian noise.  We investigated how the evolutionarily favored escape response threshold varied as a function of the parameters of their life cycle, and various properties of their predators.

Some of the results of this experiment are shown in Fig.~4. For the results in the scatter plot, the simulation was repeated many times. For each run of the simulation, the parameters $(\alpha,L,m,n,R,\omega)$ were randomly selected from i.i.d. uniform distributions with the ranges specified in table~\ref{tb:ranges}. The population average threshold was recorded after each time step, and the result shown on the plot is the average over the last $10^3$ time steps. This reduces the variance of the results that stems, in part, from the variance of the children with respect to their parents, and in part from the relatively small population ($N = 2\times 10^3$) that was used in the simulation. This variance is depicted in Fig.~4.

The results of the simulation (Fig.~4) demonstrate that a population of animals whose life cycle is well-described by the automaton model in Fig.~1 will naturally evolve to display the strategy defined in Eqs.~\ref{eq:optimalfprime} and~\ref{eq:optimalex}. Much of the challenge in applying our method to real prey animals will be in accurately modeling their life cycle with a probabilistic automaton model. 

We stress that we made a specific choice of objective function for our analytic calculation, but that objective function was not available to the animals in our simulation. Had we made a different choice of objective function, our analytic calculations would have yielded different results, and those would necessarily not have been in agreement with the simulation results.

For example, choosing  longevity as an objective function, one would choose the strategy that maximizes lifetime. Given the structure of our automaton model, that strategy is clearly to flee all of the time; $\tau = -\infty$, regardless of the model parameters. That result is clearly in disagreement with our simulation results. Thus, we argue that the results of our simulation support our chosen objective function.

It has previously been conjectured~\cite{Cooper2007} that, because the correct objective function is unknown, and prey animals have uncertain information about the environment, quantitative behavioral predictions are impossible. We have addressed both of these issues: the correct objective function, while hard to compute for real prey animals (much information is required to correctly estimate $r$), is known, and we have explicitly incorporated the effects of imperfect information into our decision model. 

We conclude that, given sufficiently accurate and detailed (probabilistic) information about the life cycle of an animal, it may be possible (although difficult) to make quantitative behavioral predictions.

\section*{Discussion and Conclusions}

We have found that a prey animal's uncertainty about threats in its environment has a profound effect on the optimal escape strategy. Moreover, computer simulations of the evolution of populations of animals subject to predation demonstrate that the objective function we assumed for our analytic calculations is, indeed, optimized by selection pressure.

Interesting work has modeled the learning process in the presence of uncertainty, in the context of optimal decision making~\cite{rao}. Our results focus instead on instinctual responses, and do not explicitly incorporate learning over the lifetime of the animal. Clearly, in the real world, both innate and learned behaviors are important. We leave the issue of combining these two response types for future work. 

Whereas much previous work\cite{YdenDill, Broom2005, Cooper2007, Cooper2010, Cooper2006, Blumstein} has involved determining flight initiation distances, our model does not do so explicitly. In our model, the animal simply flees when the possibility of danger exceeds some threshold, the value of which is determined by the level of uncertainty, and, when that uncertainty is not small, by economic factors. However, we do have the ability to infer how such a strategy might vary when assessing threats at different distances. 

We expect that nearby threats will be more conspicuous: $|\omega|$ should be a decreasing function of distance. Thus, the economic factors are more important for potential threats at large distances compared to small. Consequently, those economic factors that make the strategy more timid (lower threshold) will increase the flight initiation distance -- they make the optimal threshold lower at large distances, but do not affect the small distance threshold as strongly. 

Similarly, when an intruder initiates its approach from further afield, the prey animal has more time to gain information about it.. Thus, at a greater distance, the animal can correctly assess the threat, leading to a larger flight initiation distance, as observed in real prey animals\cite{Blumstein}. 

Finally, when an odor or shape is presented to the animal that is associated with common predators, the score of the intruder will be far from the mean of the ``no threat" distribution. Thus, defensive behavior is likely to be trigered.

Our decision-theoretic model for prey escape strategy can thus account for several observed behaviors~\cite{Blumstein, Cooper2006, Apfelbach} in a natural way. Indeed, to the best of our knowledge, ours is the first model to account (qualitatively) for the influence of all of these factors on escape decisions.

 We propose that approaches based on optimal performance in the face of imperfect information are likely to be useful for studying further aspects of escape decisions in prey animals, as they have been in other areas of biology such as mate selection\cite{Benton,Luttbeg}, house-hunting\cite{Marshall}, cellular-level decision processes\cite{Perkins}, and chemotaxis\cite{adler_74, bialek_setayeshgar_2005}. Our approach is easily generalized to include other areas where decisions must be made with imperfect information and the costs of type I and II errors are unequal (when the costs of both
error type are equal, there are simpler tools, such as the
Neyman-Pearson lemma~\cite{Neyman}, for assessing the optimal strategy). Immunology is one such area: excessive type I errors by macrophages result in infection of the host, while excessive type II errors result in autoimmune disorders\cite{Morris}.

\section*{Acknowledgments}
The authors are grateful to the William J. Fulbright Foundation and the University of California for funding this work.

\bibliography{biblio.bib}

\newpage

\section*{Figure Legends}

\begin{figure}[ht!]
\begin{center}
\includegraphics[width=6.0in]{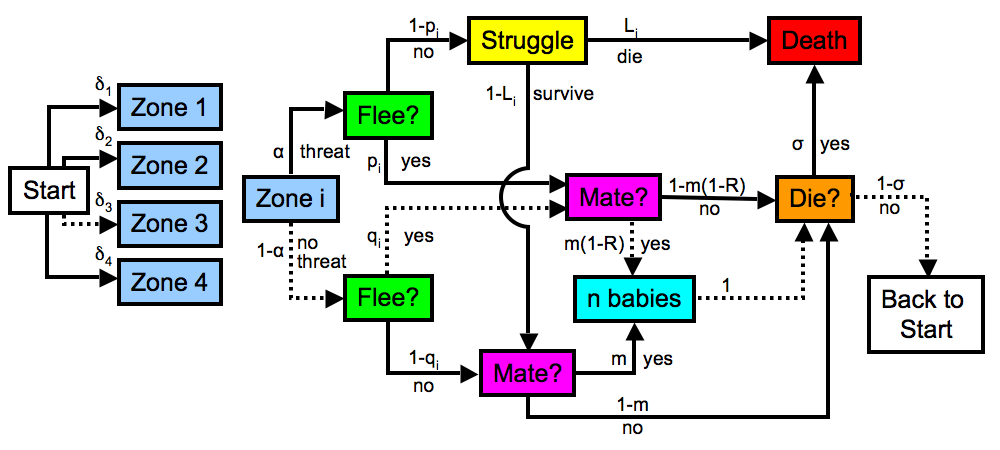}
\end{center}
 \caption{ {\bf Probabilistic automaton model of the life cycle of a prey animal.} At each time step, every animal begins in the state ``start," and follows a complete path, ending either back at the start, or in death. Each arrow is labeled with the conditional probability that the given event occurs (die, survive, etc.), once the animal reaches the box at the tail of that arrow. The animal spots a potential threat in zone $i$ with probability $\delta_i$.  Four zones (groupings by predator-prey distance) are shown in the diagram. The threat is real with probability $\alpha$. If there is a threat the prey animal flees with probability $p_i$. Those animals that do not flee are killed with probability $L_i$, while those that do flee always escape. The animals that neither flee nor die mate with probability $m$, producing $n$ progeny. The animals that do flee suffer a reduced mating rate of $m(1-R)$. In order to keep the population stable, some randomly selected animals are killed at the end of the time step with probability $\sigma$. The probability of any path is obtained by multiplying the conditional probabilities of each subsequent step. A sample path is illustrated by the dashed arrows in the diagram: The animal spots a potential threat in zone 3. This is not a real threat, but, with its imperfect information, the animal incorrectly decides to flee. It then mates, producing $n$ progeny, which are added to the population for the next time step. The animal does not succumb to disease, starvation, or other non-predation-related causes of death, and lives on to the next time step. The probability of this path is $\delta_3(1-\alpha)q_im(1-R)(1-\sigma)$.}  
 \label{fg:markov}
\end{figure}

\newpage

\begin{figure}[ht!]
\begin{center}
 \includegraphics[width=3.4in]{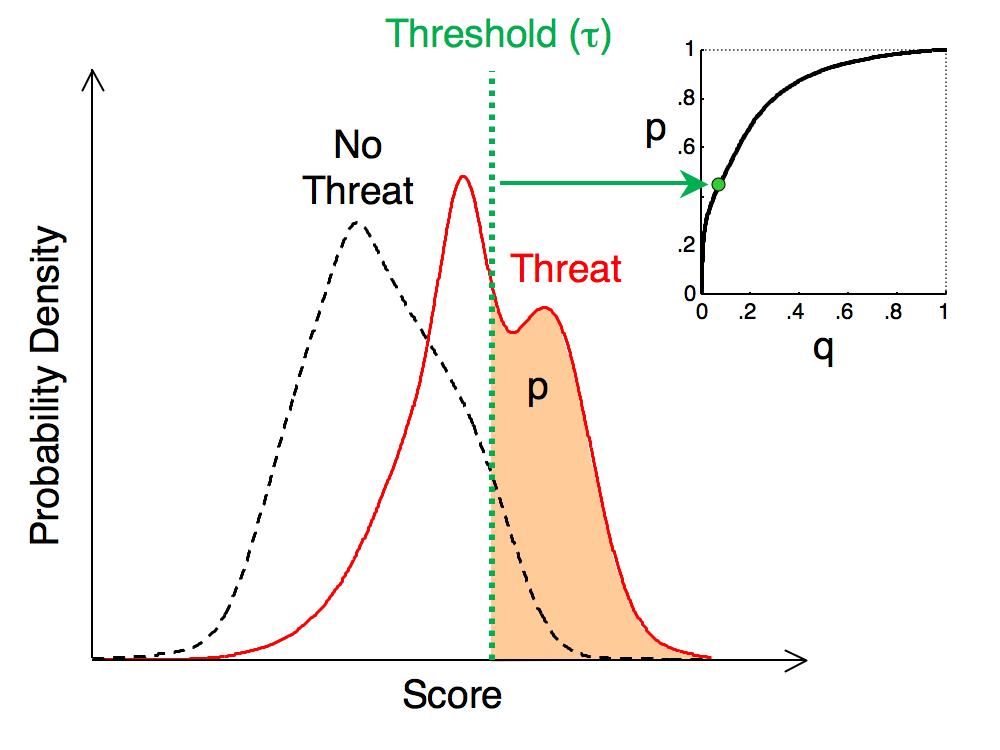} 
\end{center}
  \caption{ {\bf Connection between ROC curves ($p$ vs. $q$) and probability distributions.} The probability distributions of some ``score" parameter, conditioned on the presence (solid curve) or absence (dashed curve) of a threat are shown. A possible interpretation for this score is that it is the output of a neural network that assesses all of the information available to the animal, in an attempt to infer the danger of a given object. One possible decision threshold, $\tau$, is indicated, whereby the animal decides to flee from objects with scores above the threshold, and not to flee from those with scores below the threshold. By varying the threshold, the animal can alter the correct detection probability $p$, indicated by the area under the ``threat" distribution (solid curve) to the right of the threshold; at the same time, varying the threshold also affects the false positive probability $q$, given by the area under the ``no threat" distribution (dashed curve) to the right of the threshold. The same threshold determines both $p$ and $q$; they are related by the ROC curve $p = f(q)$ (inset).}  
 \label{fg:distributions}
\end{figure}

\newpage

\begin{figure}[ht!]
\begin{center}
  \includegraphics[width=3.4in]{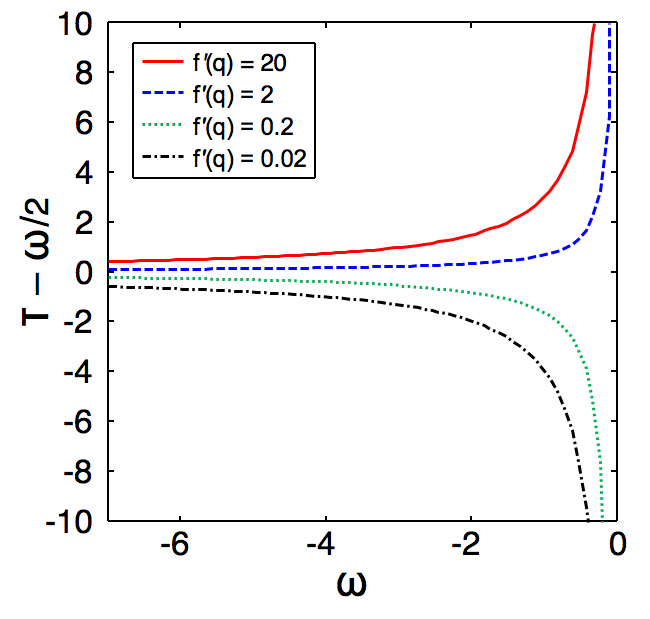}
\end{center}
  \caption{ {\bf The importance of economic factors in
decision making increases with rising uncertainty about the
environment}. The departure of the optimal decision threshold ($\tau$) from a maximum likelihood estimator (described by $\tau = \omega/2$) is shown as a function of $\omega$, the displacement between the means of the score distributions for threats and non-threats. The result is shown for several different values of $f'(q)$, which contains all the economic factors, and quantifies how bold (large values) or timid (small values) the strategy is (see text). For large $| \omega |$, threats are easily identified, and the strategies all converge to a maximum likelihood decision strategy: flee if and only if danger is more likely than not. As the uncertainty increases (small $| \omega |$), the strategies diverge in a manner dictated by the economic factors.}  
 \label{fg:uncertainty}
\end{figure}

\newpage

\begin{figure}[ht!]
\begin{center}
\includegraphics[width=6.0in]{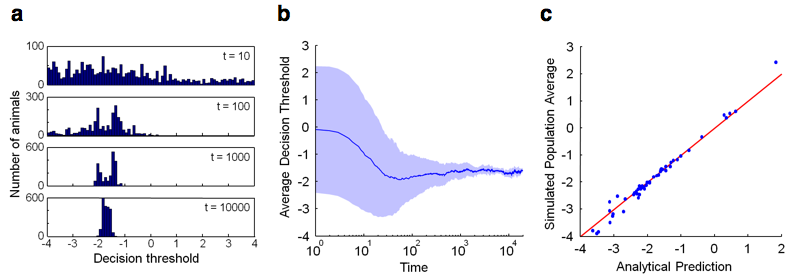}
\end{center}
\caption{ {\bf Computer simulation confirms that our objective function, $r = N^{-1}dN/dt$, is indeed maximized by selection pressure.}  {\bf a}, Time evolution of the distribution of escape response thresholds across a population of simulated prey animals. The time (in units of time steps) at which the histogram was measured is indicated on each histogram. The population was initialized at time $t=0$ with a uniform distribution of escape strategies. The model parameters for this simulation were $(\alpha,L,m,n,R,\omega) = (0.15,0.8,0.02,4,0.5,-2.0)$, and the threshold of each progeny was equal to that of its parent, plus Gaussian noise with mean $0$ and standard deviation $0.01$. {\bf b}, Average escape threshold across this simulated population asymptotes to the predicted value. The shaded region extends one standard deviation above and below the average. At the end of this simulation, the population mean is $-1.605$ with standard deviation $0.08$, in good agreement with the theoretical value (Eq. 8) of $-1.645$. {\bf c}, Repeating this simulation 50 times with randomly selected parameter values shows good agreement between the analytical prediction and simulation results across the full range of parameter values tested (table~1).  Population average thresholds (after $10^4$ time steps) are plotted against the analytical prediction. The red line represents equality between the prediction and simulation.}
\label{fg:simulation}
\end{figure}

\newpage

\section*{Tables}

\begin{table}[htb!] 
  \begin{center}
    \caption{{\bf Ranges of variables used in the simulation.}
     \label{tb:ranges}}
    \begin{tabular}{ c|c } 
        \hline \hline
  Variable  	&	Range	 		\\ \hline  
$  \alpha $& [0.05, 0.6] \\
  L & [0.5,1.0] \\
  m & [0.01, 0.5] \\
  n & [3,8] \\
  R & [0.1,0.7] \\
$  \omega $& [-7.0, -0.5] \\        	
     \hline \hline
    \end{tabular}
    \end{center}
\end{table}


\end{document}